\documentclass{iaus}
\usepackage{graphicx}
\usepackage{floatflt}

\newcommand{\strom}{\mbox{Str\"omgren~}}
\newcommand{\omc}{\mbox{$\omega$ Cen~}}

\title[Metallicity distribution of \omc RGs] 
{Metallicity distribution of \omc Red Giants based on the \strom $m_1$
metallicity index}

\author[Calamida et al.]   
{A. Calamida$^{1,2}$, G. Bono$^1$, L.M. Freyhammer$^3$,
F. Grundahl$^4$, C.E. Corsi$^1$, P. B. Stetson$^5$, R. Buonanno$^2$.
M. Hilker$^6$, T. Richtler$^7$}

\affiliation{$^1$INAF-Osservatorio Astronomico di Roma, Via Frascati 33, 00040, Monte Porzio 
Catone, Italy \break email: calamida@mporzio.astro.it \\[\affilskip]
$^2$Universita' di Roma Tor Vergata, Via della Ricerca Scientifica 1,
00133 Rome, Italy \\[\affilskip]
$^3$Centre for Astrophysics, University of Central Lancashire, Preston PR1
2HE, UK \\[\affilskip]
$^4$Institute of Physics and Astronomy, Aarhus University, Ny Munkegade, 8000 Aarhus C, Denmark \\[\affilskip]
$^5$DAO, HIA-NRC, 5071 W. Saanich Road, Victoria, BC V9E~2E7, Canada \\[\affilskip]
$^6$ESO, Karl-Schwarzschild-Str. 2, D-85748 Garching bei M\"unchen, Germany  \\[\affilskip]
$^7$Universidad de Concepcion, Departamento de Fisica, Casilla 106-C, Concepcion, Chile}

\pubyear{2007}
\volume{241}
\pagerange{--}
\date{}
\jname{Stellar Populations as Building Blocks of Galaxies}
\editors{A. Vazdekis, \& R. Peletier, eds.}

\begin{document}

\maketitle

\begin{abstract}
We adopted {\it uvby\/} \strom photometry
to investigate the metallicity distribution of \omc Red Giant (RG) stars.
We provided a new empirical calibration
of the \strom $m_1\equiv(v-b)-(b-y)$ metallicity index based on cluster stars.
The new calibration has been applied to a sample of \omc RGs. The shape of
the estimated metallicity distribution is clearly asymmetric, with a sharp cut-off at low
metallicities ($[Fe/H] <$ -2.0) and a metal-rich tail up to $[Fe/H] \sim$ 0.0. 
Two main metallicity peaks have been identified, around $[Fe/H]\approx$ -1.9 and
-1.3 dex, and a metal-rich shoulder
at $\approx$ -0.2 dex.
\end{abstract}

\vspace{-0.5cm}
\section{Introduction}
The intermediate-band \strom photometric system (\strom 1966; Crawford 1975)
presents several indisputable advantages when compared
with broad-band photometric systems such as the Johnson-Cousins-Glass  
(Cousins 1976). 
The key advantages of the \strom photometric system are: {\em i)} the 
possibility to provide robust estimates of intrinsic stellar parameters 
such as the metal abundance ($m_1\equiv(v-b)-(b-y)$ index,  
Anthony-Twarog \& Twarog 2000, hereinafter ATT; Hilker 2000, hereinafter H00),
the surface gravity ($c_1\equiv(u-v)-(v-b)$ index), and the effective temperature 
($H_\beta$ index, Nissen 1988; Olsen 1988; ATT). 
Furthermore, theoretical and empirical 
evidence (Stetson 1991; Nissen 1994; Calamida et al. 2005) suggest that 
the reddening free $[c_1]\equiv c_1 - 0.2 \times (b-y)$ index is a robust reddening indicator for HB
stars hotter than 8,500 K.    
{\em ii)} Accurate \strom photometry can be adopted 
to constrain the ensemble properties of stellar populations in complex 
stellar systems like the Galactic bulge (Feltzing \& Gilmore 2000)
and the disk (Haywood 2001).  
{\em iii)} \strom photometry has been recently adopted to investigate 
the membership and the metallicity distribution of RG stars in the 
Local Group dwarf spheroidal galaxy Draco (Faria et al. 2006). 
Moreover, \strom photometry it has also been adopted to remove the 
degeneracy between age and metallicity in stellar systems hosting 
simple stellar populations (globular clusters, elliptical galaxies) and 
to investigate age and metallicity distribution of dwarf elliptical galaxies
in the Coma and Fornax clusters (Rakos \& Schombert 2005).         
On the other hand, the \strom system presents two substantial 
drawbacks: {\em i)} the $u,~v-$bands have short effective  
wavelengths, namely $\lambda_{eff}=3450,4.110$ \AA. 
As a consequence the possibility to perform accurate photometry 
with current CCD detectors is hampered by their reduced sensitivity 
in this wavelength region. 
{\em ii)} The intrinsic accuracy of the stellar parameters,  
estimated using \strom indicators, strongly depends on 
the accuracy of the absolute zero-point calibrations. This 
typically means an accuracy better than 0.03 mag. This limit 
could be easily accomplished in the photoelectric photometry 
era, but it is not trivial at all in the CCD photometry era.   
Moreover and even more importantly, current 
empirical calibrations of the \strom metallicity index are 
based either on field stars (Anthony-Twarog \& Twarog 1998) 
or on a mix of cluster and field stars (Schuster \& Nissen 1989; H00).  
However, empirical spectroscopic evidence suggest that 
field and cluster stars  present different heavy element abundance
patterns (Gratton, Sneden \& Carretta 2004). Moreover, the occurrence 
of CN and/or CH rich stars in GCs (Grundahl, Stetson, \& Andersen 2002,
hereinafter GSA02) along the RG (H00), the subgiant and the main
sequence (Stanford et al. 2004; Kayser et al. 2006) opens the
opportunity of an independent calibration of the \strom metallicity index
only based on cluster stars as originally suggested by Richtler (1989).
\vspace{-0.6cm}
  \section{Calibration of the \strom metallicity index}\label{cali}
\begin{floatingfigure}[r]{.45\textwidth}
\centering
\vspace{-0.8truecm}
\includegraphics[height=7.5cm,width=7.5cm]{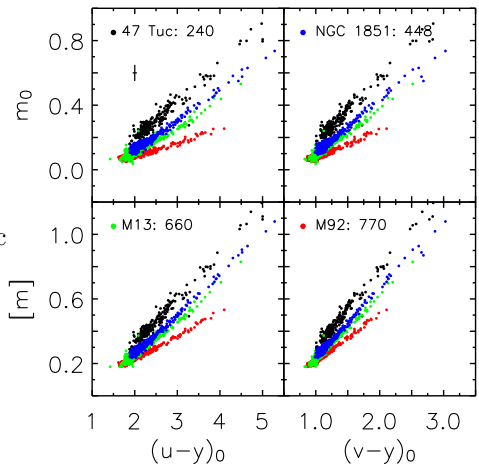}
\vspace{-0.65truecm}
\caption{Candidate RG stars for the four calibrating clusters plotted 
in different MIC planes. Dots of different colors mark 
RG stars of different clusters. The error bars in the top left panel account 
for uncertainties both in the photometry and in the reddening correction. 
The number of selected RG stars in each cluster is given. \label{fig:fig1}}
\end{floatingfigure}
In order to calibrate the metallicity index $m_1$ we selected four globular
clusters, namely M~92, M~13, NGC~1851, NGC~104, that cover a broad range in 
metallicity ($-2.2<[Fe/H]<-0.7$), are marginally affected by reddening 
($E(B-V) \le 0.04$), and for which accurate \strom photometry 
well-below the Turn-Off region is available. The photometry for these clusters 
was collected with the 2.56m~Nordic Optical Telescope
and with the Danish~1.54m Telescope. The reader interested in details of the
observations, data reduction and calibration procedures is referred to Grundahl et al. (1999, GSA02) and to Calamida et al. (2005, 2007, in preparation). 
Empirical evidence suggests that the $m_1$ versus color 
relation of RG stars presents a linear trend and a good sensitivity 
to iron abundance (SN88; H00). Therefore, we selected cluster stars from 
the tip to the base of the RGB with a photometric accuracy 
$\sigma_{u,v,b,y}\le$0.03 mag for each cluster in our 
sample. However, in order to avoid subtle systematic uncertainties in the empirical
calibrations, current cluster RG stars need to be cleaned from the
contamination of field stars.
To accomplish this goal we decided to use optical-NIR color planes
to split cluster and field stars. We cross-identified stars
in common with our \strom catalogs and the
Near-Infrared Two Micron All Sky Survey (2MASS) catalog
(Skrutskie et al. 2006). We found that the best optical-NIR color-color plane to properly
identify field and cluster stars is the $u-J,\ b-H$. All cluster samples were
reduced by $\approx$ 40\% after the color-color plane selection.
The reader interested in details concerning the cleaning procedures adopted is referred to
Calamida et al. (2007).
We derived new Metallicity-Index-Color (MIC) empirical relations that correlate
the iron abundance of RG stars to their metallicity ($m_1$) and color ($CIs$) indices.
Together with the unreddened $m_1$ index ($m_{10}$, hereinafter $m_0$),
we also derived independent empirical MIC relations for the 
reddening free parameter $[m] = m_1\, +\, 0.3\times(b-y)$. This \strom 
index was adopted to overcome deceptive uncertainties in clusters 
affected by differential reddening. 
Fig. 1 shows the cleaned samples of candidate RGs for the four calibrating 
clusters in four different unreddened MIC planes. We assumed 
$E(b-y)= 0.70\times E(B-V)$, $E(v-y)=1.33\times E(B-V)$, $E(u-y)= 1.84\times E(B-V)$,
adopting the reddening law from Cardelli et al. (1989) and $R_V=3.1$. 
Reddening values for the selected clusters are from Harris (2003) 
and Schlegel et al. (1998).
The error bars plotted in the top left panel display the photometric 
error budget for the unreddened color indices 
($\sigma (u-y)_0 \le 0.05$, $\sigma (v-y)_0 \le 0.045$ mag) and for the 
metallicity indices ($\sigma (m_0/[m]) \le 0.05$ mag). 
Data plotted in Fig. 1 show two compelling empirical 
evidence: {\em i)} the $m_0/[m], CI_0$ relations are linear over a 
broad color range, namely $1.5\lesssim (u-y)_0 \lesssim 5.0$ and 
$0.85\lesssim (v-y)_0\lesssim 3.0$; {\em ii)} the metallicity indices 
are well correlated with the cluster metal abundance, and indeed the 
four calibrating clusters present sharp and well-defined slopes. 
Hence, we applied a multilinear regression fit, by adopting 
cluster metallicities, to estimate the coefficients of 
the four MIC relations. Note that the metallicities adopted in 
the fit are in the Zinn \& West (1984) scale and are based on the Calcium triplet 
measurements provided by Rutledge et al. (1997). 
\section{Metallicity distribution of \omc RGs}\label{sec:met}
We applied the new MIC relations to estimate \omc RGs metal abundances.
In order to verify the reliability of these estimates,
we cross-correlated this sample with the high-resolution
spectroscopy of 40 ROA stars by Norris et al. (1995), finding 26 common RGs.
The difference between photometric metallicities estimated adopting the 
$[m],\ (v-y)_0$ MIC relation and the
spectroscopic measurements are plotted versus spectroscopic abundances in
Fig. 2. The error bars plotted in the figure accounts for photometric and  
reddening uncertainties and spectroscopic measurement errors.
The agreement between photometric and spectroscopic abundances
is very good, with $\Delta [Fe/H] = [Fe/H]_{phot} - [Fe/H]_{spec} \lesssim$ 0.3 dex
for most of the stars, and the mean of the residuals is $\sim 0.02$ dex.
However, seven RGs (marked with an asterisk)
show systematic higher photometric abundances. We found that these
stars are labeled as CN-strong stars in the list of Norris.
Without taking into account these peculiar stars the dispersion of the residuals
is $\sigma \lesssim$ 0.1 dex. Fig. 3 shows the metallicity distribution of selected \omc RGs, 
estimated adopting the $m_0,\ (u-y)_0$ 
MIC relation. The shape of the distribution is clearly asymmetric, with a sharp
cut-off at low metallicities ($[Fe/H]_{phot} <$ -2.0) and a metal-rich tail up to
$\approx$ 0.0 dex. 
Two main metallicity peaks can be identified 
around $[Fe/H]_{phot}\approx$ -1.9 and -1.33 dex, and
a metal-rich shoulder at $[Fe/H]_{phot}\approx$ -0.20 dex.
Fig. 3 shows the best fitting
curve to the distribution: given the non-gaussian shape
the fit has been performed adopting a convolution of three gaussians. 
The three gaussian curves corresponding to different stellar groups are 
overplotted to the distribution.
The figure shows that the dispersion of the metal-poor gaussian
is $\approx$ 0.2 dex, accounting for the photometric and the reddening
uncertainties. Moreover, the lack of a resolved metal-poor tail in 
our \omc metallicity distribution suggests that these stars could have formed
in a chemically homogeneous environment, as already found by
Norris et al. (1996) based on their spectroscopic RGs metallicity distribution.
However, before we can reach firm conclusions concerning our metallicity
distribution we still have to constrain on a quantitative basis the dependency of
the photometric abundance estimates on the occurrency of CN and CH rich stars in \omc.
\begin{figure}[h!]
\centering
\vspace{-0.4truecm}
\begin{minipage}[l]{.45\textwidth}
\centering
\includegraphics[height=6cm,width=6.2cm]{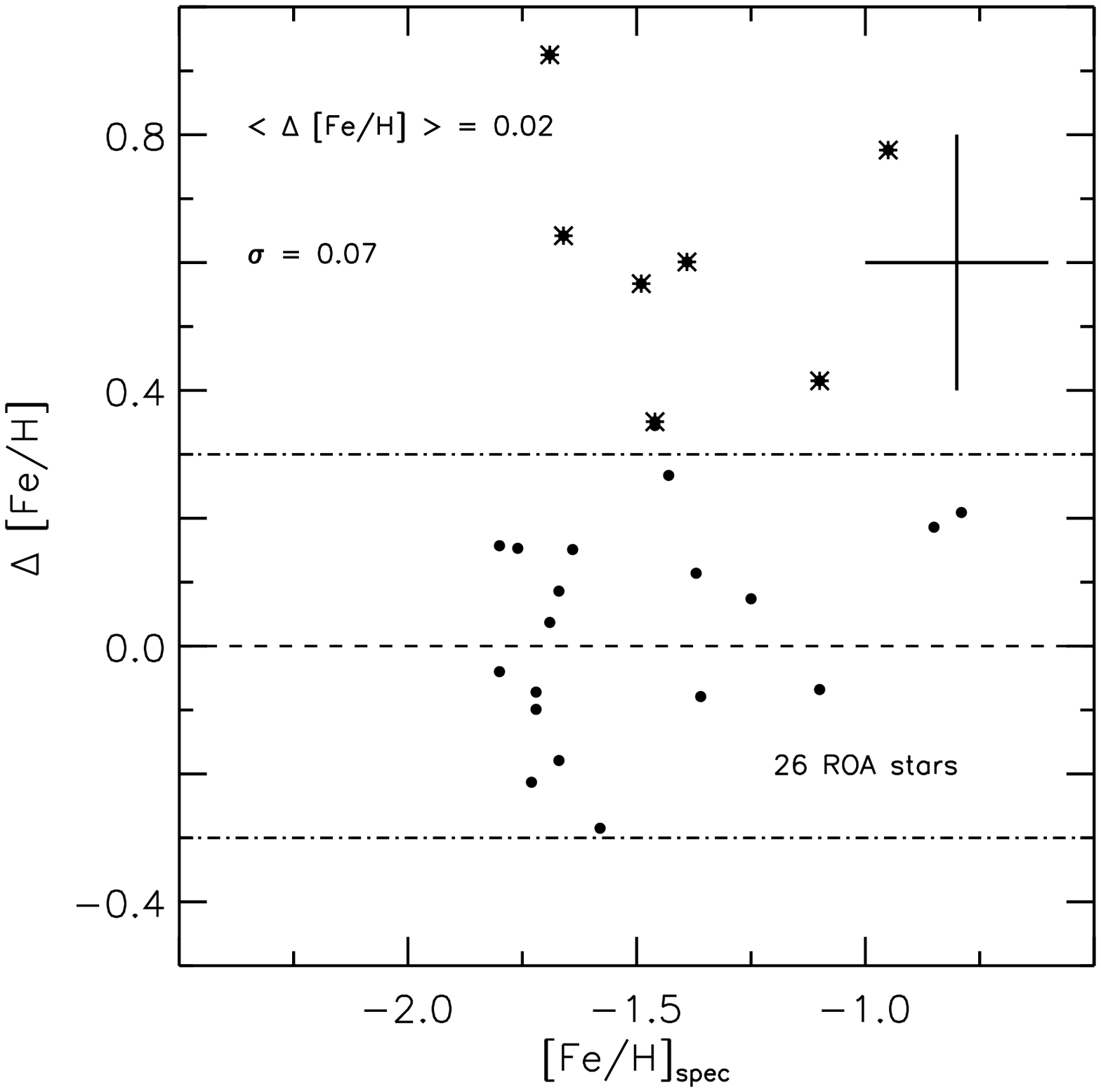}
\hspace{35mm}
\vspace{-0.55truecm}
\caption{Difference between photometric
and spectroscopic metallicities plotted versus $[Fe/H]_{spec}$ for 26 \omc ROA stars 
from the list of Norris. Asterisks mark
CN-strong stars according to Norris. The error bars
accounts for both photometric estimate and spectroscopic measurement
errors.\label{fig:fig2}}
\end{minipage}
\hspace{8mm}
\vspace{-0.25truecm}
\begin{minipage}[r]{.45\textwidth}
\centering
\includegraphics[height=6cm,width=6.2cm]{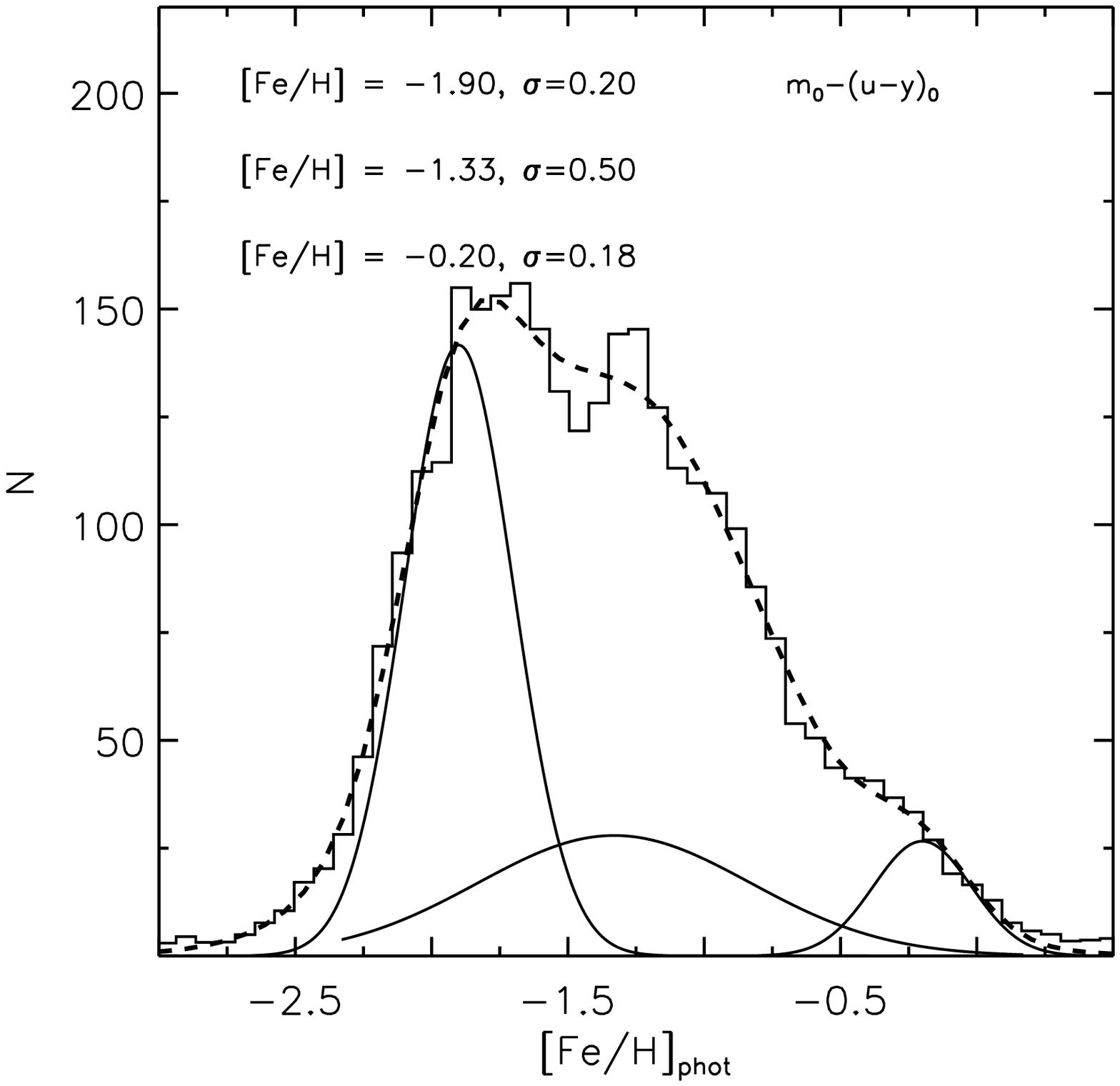}
\vspace{-0.6truecm}
\caption{\omc RGs metallicity
distribution estimated adopting the $m_0,\ (u-y)_0$ empirical MIC relation.
The dashed line shows the best fitting curve obtained with the convolution of three
gaussians. The solid lines show the three individual gaussians. The 
$[Fe/H]_{phot}$ peak values and relative dispersions are also labeled.\label{fig:fig3}}
\end{minipage}
\end{figure}

\end{document}